\documentstyle[12pt]{article}

\def\fun#1#2{\lower3.6pt\vbox{\baselineskip0pt\lineskip.9pt
        \ialign{$\mathsurround=0pt#1\hfill##\hfil$\crcr#2\crcr\sim\crcr}}}

\renewcommand\({\left(}
\renewcommand\){\right)}

\newcommand\eq[1]{Eq.~(\ref{#1})}
\newcommand\eqs[2]{Eqs.~(\ref{#1}) and (\ref{#2})}

\newcommand\eqst[2]{Eqs.~(\ref{#1})--(\ref{#2})}

\newcommand\ee{\end{equation}}
\newcommand\be{\begin{equation}}
\newcommand\eea{\end{eqnarray}}
\newcommand\bea{\begin{eqnarray}}

\newcommand\TeV{\,\mbox{TeV}}
\newcommand\GeV{\,\mbox{GeV}}
\newcommand\MeV{\,\mbox{MeV}}

\newcommand\mpl{M_{\rm P}}

\newcommand\mpsis{|m_\chi^2|}
\newcommand\etapsi{\eta_\chi}
\newcommand\luv{\Lambda_{\rm UV}}

\newcommand\lsim{\mathrel{\rlap{\lower4pt\hbox{\hskip1pt$\sim$}}
    \raise1pt\hbox{$<$}}}
\newcommand\gsim{\mathrel{\rlap{\lower4pt\hbox{\hskip1pt$\sim$}}
    \raise1pt\hbox{$>$}}}

\def\dslash{\not{\hbox{\kern-2pt $\partial$}}}
\def\Dslash{\not{\hbox{\kern-4pt $D$}}}
\def\Oslash{\not{\hbox{\kern-4pt $O$}}}
\def\Qslash{\not{\hbox{\kern-4pt $Q$}}}
\def\pslash{\not{\hbox{\kern-2.3pt $p$}}}
\def\kslash{\not{\hbox{\kern-2.3pt $k$}}}
\def\qslash{\not{\hbox{\kern-2.3pt $q$}}}

 \newtoks\slashfraction
 \slashfraction={.13}
 \def\slash#1{\setbox0\hbox{$ #1 $}
 \setbox0\hbox to \the\slashfraction\wd0{\hss \box0}/\box0 }
 
\def\ee{\end{equation}}
\def\be{\begin{equation}}

\newcommand\sub[1]{_{\rm #1}}

\newcommand\phicob{\phi\sub{COBE}}
\newcommand\delmult{\Delta V_{\chi\tilde\chi{\rm f}}}

\begin{document}

\begin{flushright}
LANCS-TH/9915
\\hep-ph/9908219\\
(April 1999)
\end{flushright}
\begin{center}
{\Large \bf Constraints on TeV-scale hybrid inflation
and comments on non-hybrid alternatives}

\vspace{.3in}
{\large\bf  David H.~Lyth}

\vspace{.4 cm}
{\em Department of Physics,\\
Lancaster University,\\
Lancaster LA1 4YB.~~~U.~K.\footnote
{A subset of the hybrid inflation constraints appeared
in the unpublished note hep-ph/9904371}
}
\vspace{.4cm}
\end{center}

\begin{abstract}
During hybrid inflation, the slowly-rolling inflaton field
has a significant coupling to the trigger field which is 
responsible for most of the potential.
Barring a fine-tuned accidental cancellation, this
coupling induces a minimal one-loop contribution to the 
inflaton potential. The requirement that this contribution be not 
too large constrains a wide class of 
hybrid inflation models.
Assuming that the inflaton perturbation generates structure in the 
Universe, the inflaton field and/or the
trigger field after inflation have to be bigger than $10^9\GeV$.
This and other results make hybrid inflation at or below the TeV
scale problematical. (There is no problem with hybrid inflation at
the high energy scales normally considered.)
`New' and thermal inflation seem to be viable alternatives for inflation
at or below the TeV scale, 
including the case that quantum gravity is at the TeV scale.
In any case, supersymmetry is needed required during inflation, 
in order to protect a scalar mass. 
\end{abstract}

\paragraph{Introduction} 
Hybrid inflation, where some `trigger' field $\chi$
responsible for the
bulk of the potential is different from the slowly-rolling inflaton
field $\phi$, has proved a very useful paradigm which may in the end
turn out to be the one chosen by Nature.

The original model \cite{andhyb} worked with a tree-level potential,
the slope of the inflaton 
potential being given by the mass term $\frac12m^2\phi^2$. 
The potential was soon shown to be derivable from 
spontaneously broken global
supersymmetry (susy), with either the $F$ term 
\cite{cllsw,ewansg} or the $D$ term \cite{ewansg} dominating.\footnote
{In these models the inflaton mass vanishes at
the level of global supersymmetry,
but supergravity corrections can give
a suitable mass which indeed tends to be somewhat too large in the 
$F$ term model.}
For each type of model, the one-loop correction
to the inflaton potential
coming from the trigger field and its superpartners 
(trigger supermultiplet) was evaluated
\cite{giaf,giad}.
Later, realizations of the tree-level model 
in the context of
softly broken supersymmetry were given \cite{rsg,laura}, leading to a 
one-loop correction of a different form 
\cite{ewanloop,laura}.
In order to have a more attractive model in the context of 
supergravity, the dominant loop correction in this case is supposed to
come from some gauge supermultiplet, not from
the trigger supermultiplet

In this note I go beyond specific models. 
Barring accidental cancellations, the loop
correction from the trigger supermultiplet 
cannot be less than it is in the case of the models with spontaneously 
broken global supersymmetry.
Still barring accidental cancellations, this
places a lower bound on the derivative of the inflaton potential,
which in turn limits the allowed region of parameter
space for a wide class of hybrid inflation models. 
Assuming that the
inflaton fluctuation is the origin of structure in the Universe,
we find the bound
\be
M^4\phicob \gsim \( 10^9\GeV \)^5\,, 
\label{advert}
\ee
where $\phicob$ is the inflaton field when scales explored by
COBE leave the horizon, and 
$M\equiv \langle\chi\rangle$ is 
the vacuum expectation value of the trigger field.
Whether or not the inflaton fluctuation is the origin of structure,
we find the bound
\be
M^2\phi_N \gsim \(10^4\GeV \)^3 \sqrt N\,,
\label{advert2}
\ee
where $\phi_N$ is the inflaton field $N$ $e$-folds before the end of 
slow-roll inflation.
Implications of these results for TeV-scale inflation will be 
considered at the end of the paper.

Throughout the paper, `hybrid inflation' is taken to mean a model
in which the trigger field is fixed during inflation, with the
inflaton field moving towards the origin. This excludes
`inverted' hybrid inflation in which the inflaton is moving away from
origin 
\cite{burt,ourinvert,steve} and `mutated' hybrid inflation 
\cite{ewanmut} in which the
trigger field has a time-dependent value adjusted to
minimize the potential. Only hybrid inflation in the narrow sense is 
treated here.

\paragraph{Basics}
Let us summarize the basics of inflation model building, as given in say
\cite{treview}.
As usual, $\mpl\equiv (8\pi G)^{-1/2}
=2.4\times 10^{18}\GeV$ is the Planck scale while $H$ is the Hubble
parameter. Our Universe is assumed to be flat so that after it leaves
the horizon $3\mpl^2H^2\simeq
\rho$ where $\rho$ is the energy density.
An overdot denotes
differentiation with respect to time and the prime differentiation with
respect to the inflaton field $\phi$.
              
We are concerned with the
slow-roll paradigm of inflation  
in which the field equation $\ddot\phi+3H\dot\phi+V'=0$
is replaced by the slow-roll condition $3H\dot\phi\simeq-V'$, and $V
\simeq \rho$ is
almost constant on the Hubble timescale. The latter condition 
requires
the flatness condition
\be
\epsilon \ll 1 \label{flat1} \,,
\ee
and differentiating the slow-roll condition requires
another flatness condition
\be
|\eta| \ll 1 \label{flat2} 
\,,
\ee
where 
\bea
\epsilon&\equiv &\frac12\mpl^2(V'/V)^2 \,,\\
\eta&\equiv &\mpl^2 V''/V \,, \,.
\eea
$N$ $e$-folds before the end of slow-roll inflation, the field value 
$\phi_N$ is given by 
\be
N = \mpl^{-2}\int^{\phi_N}_{\phi_{\rm end}} \frac V{V'} d\phi \,,
\label{nint}
\ee
where $\phi\sub{end}$ marks the end of slow-roll inflation.

If the inflaton field fluctuation is responsible for structure in the 
Universe, the COBE measurement of the cosmic microwave background anisotropy
requires
\be
\mpl^{-3} V^{3/2}/V' = 5.3\times 10^{-4} \,.
\label{cobenorm}
\ee
This equation applies at the epoch when the distance scale
explored by COBE (say $H_0^{-1}/10$) leaves the horizon,
some number $N\sub{COBE}<60$ $e$-folds before the end 
of slow-roll inflation.

\paragraph{Hybrid inflation}
The original tree-level hybrid inflation model \cite{andhyb}
is defined by
\be
V(\phi,\chi)= V_0 + \Delta V(\phi) 
-\frac12\mpsis\chi^2 +\frac12{\lambda'}\chi^2\phi^2
+ \frac14\lambda\chi^4 \,,
\label{fullpot}
\ee
where
\be
\Delta V(\phi) = \frac12m^2\phi^2 \,.
\label{mterm}
\ee
Inflation takes place in the regime $\phi^2>\phi\sub c^2$, where
\be
\phi\sub c \equiv |m_\chi|/\sqrt{\lambda'} \,.
\label{phic}
\ee
In this regime, $\chi$ vanishes 
and the inflaton potential is
\be
V=V_0+ \Delta V(\phi) \,. \label{vofphi2}
\ee
The constant term $V_0$ is assumed to dominate during inflation.

The last term of \eq{fullpot} serves only to determine the vacuum 
expectation value (vev) of $\chi$, achieved when $\phi$ falls below
$\phi\sub c$. Using
that fact that $V_0$ vanishes in the vacuum, one learns that
the vev is
\be
\langle\chi\rangle\equiv
M= 2 V_0^{1/2}/|m_\chi| \,, 
\label{mofv} \ee
and that
\be
\lambda=\frac{4V_0}{M^4} = \frac{m_\chi^4}{4V_0} \,.
\label{lamofv}
\ee

{}From \eq{mterm},
\be
\eta = \frac{m^2 \mpl^2}{V_0} \,.
\ee
It is useful also to define
\be
\etapsi \equiv \frac{\mpsis\mpl^2}{V_0} =\frac{4\mpl^2}{M^2}
\label{etapsiapp}
\,.
\ee
A prompt end to inflation at
$\phi\sub c$ requires 
\be
\etapsi \gsim  1 \,.
\ee

As already mentioned, alternative models of hybrid inflation have
been proposed where $\Delta V$ is dominated by a loop correction instead
of by the mass term \eq{mterm}. 
One might also allow significant tree-level terms $\propto \phi^p$
with $p>2$. In any case, once
$\Delta V$ is specified,
$\phicob$ is determined by \eq{nint} with 
\be
\phi\sub{end}=\max\{\phi\sub c,\phi\sub{fast}\} \,,
\ee
where $\phi\sub{fast}$ is the field value when one of \eqs{flat1}{flat2}
fail.
Assuming that there are no fine-tuned accidental cancellations, but
{\em without} specifying the precise form of $\Delta V$,
we argue first that the magnitude of the loop correction
to $\Delta V'$, and hence of $\Delta V'$ itself, cannot be much less than
$m_\chi^4/(16\pi^2\phi)$. Inserting this into \eq{cobenorm} 
gives the advertised bound \eq{advert}, and inserting it into
\eq{nint} gives \eq{advert2}.

We go on to show that the same bounds hold for even more
general hybrid inflation potentials, and discuss their implication for 
recent proposals concerning TeV-scale inflation. Further bounds
are obtained under additional assumptions.

\paragraph{The one-loop correction}
The one-loop correction is \cite{treview}
\be
\Delta V\sub{loop}(\phi)=\frac{Q^2}{32\pi^2}\:{\rm Str}{\cal M}^2(\phi)+
\frac{1}{64\pi^2}\:{\rm Str}\left[{\cal M}^4(\phi)
\left({\rm ln}\frac{{\cal M}^2(\phi)}{Q^2}-\frac{3}{2}\right)\right]\,.
\label{pot}
\ee
Here, ${\cal M}^2(\phi)$ is the field dependent
mass-squared matrix for the particles contributing to the loop
correction. These particles will in general have spins $j=0,1/2$
or $1$, and the supertrace is defined as
\be
{\rm Str}\:A=\sum_j(-1)^{2j}(1+2j)\:{\rm Tr}\:A_j
\,,
\label{strc}
\ee
Here, $A$ denotes either ${\cal M}^2$ or the square bracket,
and $A_j$ is the ordinary trace for particles of
spin $j$.

The quantity $Q$ is the renormalization scale at which the parameters of the 
tree-level potential should be evaluated. Its choice is arbitrary,
and if all loop corrections were included,
the total potential would be independent
of $Q$ by virtue of the Renormalization Group Equations (RGEs).
In any application of quantum field theory, one 
should choose $Q$ so that the total 1-loop correction is
small, hopefully justifying the neglect of the multi-loop correction.

Unless there is a fine-tuned
cancellation
the first term of \eq{pot} induces (through the RGEs)
a radiative correction to 
$m$ of order $g\sub{max}\luv$, where $\luv$ is the ultra-violet cutoff and
$g\sub{max}$ is a measure of the dominant inflaton coupling. (If this is the 
coupling to $\chi$ and its superpartners, $g\sub{max}=\sqrt{\lambda'}$.)
As is well known, global supersymmetry ensures a precise cancellation,
making the first term of \eq{pot}
independent of
$\phi$ so that it does not contribute to the slope of the 
potential and does not affect the mass. 
Supergravity corrections will spoil the cancellation to some extent
(see for example \cite{choi}) but since we are interested only 
in a lower bound on $|\Delta V\sub{loop}'|$, and are
barring accurate {\em accidental} cancellations, we ignore such 
corrections. By the same token we ignore the
$3/2$ contribution to the second term (which actually 
depends on the renormalization scheme, the $3/2$ holding in the
$\overline{DR}$ scheme) 
leaving
\be
\Delta V\sub{loop}(\phi)=
\frac{1}{64\pi^2}{\rm Str}\left({\cal M}^4(\phi)
{\rm ln}\frac{{\cal M}^2(\phi)}{Q^2}\right)\,.
\ee

The contribution of $\chi$ is
\be
\Delta V_\chi 
=\frac{1}{64\pi^2}\(m_\chi^4(\phi)
{\rm ln}\frac{m_\chi^2(\phi)}{Q^2} \)
\,,
\ee
where
\be
m_\chi^2(\phi) \equiv \({\lambda'}\phi^2-\mpsis \)
={\lambda'}(\phi^2-\phi\sub c^2)
 \,.
\ee
Making again the optimal assumption of global supersymmetry,
the total loop correction from the trigger supermultiplet is
\be
\delmult
= \frac1{32\pi^2} \(
m_\chi^4(\phi) \ln\(\frac{m_\chi(\phi)} Q \) 
+
m_{\tilde \chi}^4(\phi) \ln\(\frac{m_{\tilde\chi}(\phi)} Q \) 
-2
m\sub f^4(\phi) \ln\(\frac{m\sub f(\phi)} Q \) 
\)
\, \,,
\label{total}
\ee
where\footnote
{\eq{32} corrects Eq.~(325) of \cite{treview}.}
\bea
m_{\tilde\chi}^2(\phi) &=& {\lambda'}\phi^2 + m_{\tilde\chi}^2 \\
m\sub f(\phi) &=& \sqrt{\lambda'}\phi + m\sub f 
\label{32}
\,.
\eea
Here $\tilde\chi$ denotes the scalar partner, while f denotes
the spin-half partner. 

As will become clear, the minimal value of $|\delmult'|$ is obtained 
with $m\sub f=0$ and $m_{\tilde\chi}^2=\mpsis$. The first condition
is usually ensured by some symmetry, and the second will usually be 
satisfied if global supersymmetry during inflation is broken 
spontaneously as opposed to softly. Let us consider this case first,
and assume for the moment that
$\phi$ is significantly bigger
than $\phi\sub c$. The logs in \eq{total}
can then be taken to have the same argument
$g\phi/Q$ and one obtains
\bea
\delmult&=& \frac{m_\chi^4}{16\pi^2} \ln(\sqrt{\lambda'}\phi/Q) \label{sp1} \\
\delmult' &=& \frac{m_\chi^4}{16\pi^2 \phi}  \label{sp2} \,.
\eea

If susy is broken softly, one will still generally have $m\sub f=0$
but now $m^2_{\tilde\chi}$ can take essentially any value.
If it is positive, $\tilde\chi$ vanishes both during 
inflation and in the true vacuum, which as in the previous case 
justifies its omission from the 
tree-level potential.\footnote
{To be precise, $\tilde\chi$ vanishes during inflation 
if $m_{\tilde\chi}\gg V_0^{1/2}/\mpl$.
In the extreme opposite case $m^2_{\tilde\chi}\ll \mpsis$
the value of $\tilde\chi$ is determined by its random quantum 
fluctuation but it plays no role during inflation. In the 
intermediate case $\tilde\chi$ 
will be a component of the inflaton field contrary to our assumption 
that it has only the one component $\phi$. 
Multicomponent models, as discussed for instance in
\cite{treview}, are not treated in this note.}
If $m^2_{\tilde\chi}$ is negative, 
the omission 
of $\tilde\chi$ from the tree-level potential
is strictly justified only if $|m_{\tilde\chi}|
=|m_\chi|$, its effect then being
the trivial replacement $\chi^2\to\chi^2+\tilde\chi^2$.
However, its omission is justified in practice more generally.
Indeed, if
$|m_{\tilde\chi}|<|m_\chi|$, the field $\tilde\chi$ is held at the 
origin until the field $\chi$ is destabilized, making the former
ineffective except for a modest
increase in the value of $V_0$.\footnote
{Again we discount very small values of $|m_{\tilde\chi}|$, which would
lead to a different model of inflation.}
The
opposite case may be discounted because it is equivalent
to an interchange of the labels
$\chi$ and $\chi'$. For simplicity we assume
$m_{\tilde \chi}^2=m_\chi^2$, noting that $|\delmult'|$
will be at least as big in other cases.

Still assuming that 
$\phi$ is appreciably
bigger than $\phi\sub c$ we find in the softly broken case
\bea
\delmult &=& \frac{{\lambda'}}{8\pi^2}
m_\chi^2 \phi^2 \ln(\sqrt{\lambda'}\phi/Q) \label{s1} \\
|\delmult'| &=& \left| \frac{{\lambda'}}{8\pi^2}m_\chi^2  \phi
\(2\ln(\sqrt{\lambda'}\phi/Q) + 1 \) \right| \sim \left|\frac{\lambda'}{8\pi^2}
m_\chi^2 \phi \right| \label{s2} 
\,.
\eea
The final equality is valid if $Q$ is chosen to make
$\ln (\sqrt{\lambda'}\phi/Q)$ (and therefore $\delmult$)
vanish for the value of $\phi$ under consideration.

Any
choice making $|\ln(\sqrt{\lambda'}\phi/Q)|\sim 1$
would be equally valid in that it would still justify
the neglect of higher loop corrections, and one could actually choose
$\delmult'=0$ at (say) the COBE scale. But as already remarked, the 
total potential (tree level plus loop correction) is independent of $Q$
by virtue of the RGE's. 
We shall now argue that the right hand side of \eq{s2} provides
a lower limit on the slope of the total potential.
The one-loop correction will be typically be
valid in an interval over which $\phi$
varies by a factor of a few, if $Q$ is set equal to some
$\sqrt{\lambda'}\phi$ in the range. 
At the one-loop level,
$\widetilde {\Delta V}_{\chi\tilde\chi{\rm f}}\equiv \frac12m^2(Q)\phi^2
+\Delta V_{\chi\tilde\chi{\rm f}}(Q,\phi)$ is 
approximately independent
of $Q$ by virtue of the one-loop RGE for $m^2$.
One can write \cite{withlaura}
 $\widetilde{\Delta V}_{\chi\tilde\chi{\rm f}}'=
(\lambda'/4\pi^2)m_\chi^2 \phi\ln(\phi/\phi_0)
$, where $\phi_0$ is some constant.
In general this makes $\widetilde{\Delta V}_{\chi\tilde\chi{\rm f}}'$ 
of the advertised magnitude \eq{s2}. If $\phi_0$ is within the range of 
validity of this expression, $\widetilde{\Delta V}_{\chi\tilde\chi{\rm f}}'$
 vanishes at $\phi=\phi_0$
and will be reduced by a factor $|1-\phi/\phi_0|$
if $\phi$ is very close to $\phi_0$.
Let us see what happens in that case.
For hybrid inflation $\phi_0$ should correspond to a maximum,
with $\phi\sub{end}<\phi<\phi_0$.
{}From \eq{nint} we learn that
$(1-\phi\sub{end}/\phi_0) = e^{|\eta| N} (1 -\phi_N/\phi_0)$,
where $\eta\equiv \mpl^2V''/V$ is evaluated at $\phi_0$.
(We assume that the expression for 
$\widetilde{\Delta V}_{\chi\tilde\chi{\rm f}}$
is valid down to $\phi\sub{end}$ which is good enough for a
crude order of magnitude estimate \cite{withlaura}.)
We 
are considering the case that $\phi_N/\phi_0=1$ with extreme accuracy;
but then, the flatness condition $|\eta|\ll 1$ requires that in the
physically relevant range $N\lsim N\sub{COBE}<60$, $\phi\sub{end}/\phi_0$ is
also equal to 1 with extreme accuracy. 
As always, we
assume that such accidental fine-tuning does not occur.
The conclusion is that indeed the right hand side of
\eq{s2} will provide a lower bound on the
slope of $\widetilde{\Delta V}_{\chi\tilde\chi{\rm f}}$.
         
Finally, consider the case that 
$\phi$ is very close to $\phi\sub c$. 
The loop contribution from $\chi$ is very small, but those of its 
superpartners are unsuppressed. Focussing say on the contribution of
$\tilde\chi$, it is easy to check that with (say)
$Q$ chosen to make $\delmult=0$ one finds a 
a result for
$|\delmult'|$ at least as big as \eq{sp2}.\footnote
{To be precise, the slope of the renormalization-group-improved
tree-level potential will be at least as big as the right hand side
of \eq{sp2}. As in the previous case, 
one could actually choose $Q$ to make $\delmult'$
vanish at a given value of $\phi$.}

\paragraph{Constraints on hybrid inflation models}
Discounting as always the possibility of
a precise 
accidental cancellation, one will have
\be
|\Delta V'| \gsim |\delmult'|
\,.
\label{dnv}
\ee
Using the minimal estimate \eq{sp2}, the COBE normalization
\eq{cobenorm} with \eq{etapsiapp} then gives 
\be
M^4 \phi\sub{COBE}
\gsim 5.4\times 10^{-5} \mpl^3 V_0^{1/2} 
= \( 10^{9}\GeV \)^5\( \frac{V_0}{1\MeV} \)^\frac12 \,.
\label{fourth}
\ee
To obtain the advertised final inequality \eq{advert},
we set $V_0^{1/4}\gsim 1\MeV$,
the smallest possible value since reheating must occur before
nucleosynthesis.

In the case of soft susy breaking \eq{s2} provides a stronger lower bound
on $\Delta V'$, which with \eq{cobenorm} gives
\be
5\times 10^{-4} \mpl^3 \lsim \frac{V_0^{3/2}}{\lambda' \mpsis \phi}
\,.
\label{soft1}
\ee
Using $\phi\sub{COBE}>\phi\sub c$ gives a bound on $M$ alone,
\be
M^3 \gsim 5.4\times 10^{-5} \sqrt{\lambda'} \mpl^3 \,.
\label{soft2}
\ee
This bound is less interesting than \eq{fourth} because it 
is evaded by the case of spontaneous symmetry breaking, and because it
can be made arbitrarily weak by lowering $\lambda'$.
Although it may not be very natural, there does not seem to be any
bar to taking $\lambda'$ very small. In particular, reheating could
take place through the coupling of the inflaton to some
field other than the trigger field.

We can obtain further results if the original model $\Delta
V\simeq \frac12m^2\phi^2$ is supposed to be valid.
In this case,
\be
\phicob^2 =e^{2\eta N} \phi\sub c^2 \simeq \phi\sub c^2 
\,.
\label{phi}
\ee
(The approximate equality corresponds to $e^{2\eta N}\sim 1$
which is good enough for order-of-magnitude estimates.)
The COBE normalization \eq{cobenorm}
becomes
\cite{treview}
\bea
{\lambda'} &=& 2.8\times 10^{-7} e^{2\eta N} \eta^2 \eta_\chi \\
&\simeq& 3 \times 10^{-7} \eta^2 \eta_\chi \,.
\label{cobenorm2}
\eea
For this original model to be valid
one needs $\delmult'\ll m^2\phi$ and $\delmult''\ll m^2$,
for all $\phi\sub c \lsim
\phi\lsim \phi
\sub{COBE}$. These two constraints are about the same, and 
may be written
\be
\frac {{\lambda'}}{32\pi^2} \frac{\etapsi} \eta \ll 1\,.
\label{b1}
\ee
We can use the COBE normalization \eq{cobenorm2} to eliminate any one of 
the three parameters. Eliminating $\eta$ reproduces \eq{soft2}.
Eliminating
$\eta_\chi$,
\eq{b1} becomes
\be
{\lambda'} \ll (\eta/22)^{3/2} \,.\label{first}
\ee
This is weaker than \eq{soft2} if
$M/\mpl\lsim 0.005(\eta/.025)^{1/4}$.
Finally, eliminating
${\lambda'}$ gives
\be
\eta\ll (90M/\mpl)^4 \,.\label{third} 
\ee

In all of this we considered the epoch of inflation that is supposed
to generate large scale structure, through the inflaton field 
perturbation. One can also consider \cite{late} an epoch of late
slow-roll inflation, lasting only a few $e$-folds, whose only purpose it 
to dilute unwanted relics. In that case the bounds we have considered 
disappear, but a significant bound still comes from the requirement
of $N\gsim 1$ $e$-folds of inflation. Inserting \eqs{dnv}{sp1}
 into \eq{nint} gives indeed
\be
\phi^2\gsim \frac{2\sqrt\lambda V_0^{1/2}}{\lambda'} +
\frac{\lambda N \mpl^2}{2\pi^2} \,.
\label{43}
\ee
The first term is just $\phi\sub c^2$, and is typically 
negligible. Combining this equation with \eq{lamofv} gives 
\be
M^2\phi_N \gsim \(10^4\GeV \)^3 \(\frac{V_0^{1/4}}{1\MeV} \)^2
\sqrt N\,.
\label{fifth}
\ee
Since $V_0^{1/4}>1\MeV$, this leads to the advertised bound
\eq{advert2}.

\paragraph{Nonrenormalizable terms}
The original tree-level potential \eq{fullpot} ignores 
nonrenormalizable terms. They are  of the form $\lambda_{mn}\luv^{4-m-n}\phi^m
\chi^n$,
with $m+n\geq 5$, where $\luv$ is the ultra-violet 
cutoff for the effective field theory relevant during inflation.
These terms summarize the physics which is ignored by the effective 
field theory.

Since quantum gravity certainly becomes
significant on Planck scales one must have $\luv\lsim \mpl$,
but $\luv$ will be smaller if the effective field theory breaks
before Planck scales are reached. This could happen in 
three ways. First,
a different field theory may take over,
containing 
fields that have been integrated out in the effective theory. 
Second, the scale of quantum 
gravity could be lower than $\mpl$ because there are extra dimensions 
with a large compactification radius.
Third, in the presence of a large compactification radius
field theory may give way to string theory well below the scale of
quantum gravity. The last two possibilities have received a lot of attention 
lately, with the focus on the lowest conceivable cutoff $\luv\sim \TeV$.

The coefficients $\lambda_{mn}$ are in principle determined by the
theory that takes over on scales bigger than the cutoff. 
For generic fields they will be roughly of
order 1, at least for moderate values of $n$ and $m$, without any 
cancellation between different non-renormalizable terms.\footnote
{For very large $n$ and/or $m$ one might expect factors like
$1/n!$ \cite{km}.}
In the context of string theory it is known that certain 
fields, such as the dilaton, 
exist for which this is not the case. Although one cannot rule out the 
possibility that some of these fields might be suitable for inflation,
attempts to use them for that purpose have not so far been successful
when one imposes the COBE normalization \cite{treview,giabrane}.\footnote
{One of these fields has been proposed as the trigger field
in a $D$-term inflation model \cite{halyo244} but the required 
properties have not been derived from the relevant string theory.}

Assuming that the non-renormalizable terms indeed have couplings of order 1,
both the inflaton field and the trigger field
will have to be $\lsim \luv$ in magnitude if these
terms are to be under control.\footnote
{We differ here from Kanti and Olive 
\cite{olive}, who in considering chaotic inflation
with large extra dimensions require only $\phi\lsim\mpl$.}
If \eq{fourth} holds, this
is possible only if 
\be
\luv\gsim 10^{9}\GeV\( V_0^{1/4}/1\MeV\)^{2/5} \label{luvbound}
\,.
\ee
In any case one has from \eq{fifth},
\be
\luv \gsim 10^4\GeV \(V_0^{1/4}/1\MeV \)^{2/3} \,.
\label{luvbound2}
\ee

Even with the relevant field values below $\luv$, 
one or more non-renormalizable terms could be significant. 
In particular,  either or both of the
interaction terms in \eq{fullpot} can be replaced by a
non-renormalizable term \cite{rsg}. Replacing them both gives
\be
V(\phi,\chi)= V_0 + \Delta V(\phi)
-\frac12\mpsis\chi^2 +\frac12{\lambda_n'}\luv^{2-n}\phi^n\chi^2
+ \frac14\lambda_m \luv^{4-m}\chi^m\,,
\label{fullpot2}
\ee
with $n>2$ and $m>4$.
The original model is recovered for
$n=2$ and $m=4$.
                       
Let us see what difference these replacements make.
Replacing $\frac14\lambda\chi^4$ is trivial, because \eq{lamofv}
still holds approximately which means that 
\eqst{fourth}{third} and \eq{fifth} remain valid.
The replacement of
$\frac12{\lambda'}
\phi^2\chi^2$ 
by 
$\frac12 \lambda_n'\luv^{2-n} \phi^n\chi^2$  
is less trivial. \eq{phic} is replaced by
\be
\lambda_n'\luv^{2-n} \phi\sub c^n = \mpsis 
\,.
\ee
In \eq{total} one has $m_\chi^2(\phi)=\lambda_n'\luv^{2-n}\phi^n+m_\chi^2$
and similarly for $m_{\tilde\chi}^2(\phi)$ and $m\sub f(\phi)$.
Discounting factors of order $n$, the main result
\eq{fourth} remains 
valid, and so does \eq{fifth}, while \eq{soft2} becomes
\be
\frac{M^3}{\mpl^3} \gsim 5.4\times 10^{-5} \( \frac {m_\chi}{\luv} \)^{1-
\frac 2 n} 
\( \lambda_n'\) ^{\frac 1 n} 
\,.
\label{soft22}
\ee
Finally, \eq{first} becomes
\be
(\lambda_n')^{\frac 2n} \( \frac{m_\chi}{\luv} \) ^{2-\frac 4 n} 
\ll (\eta/22)^{3/2} \,.\label{first2}
\,,
\ee
while \eq{third} is unchanged.
Requiring $M\lsim \luv$ in 
\eq{soft22} (with $\lambda_n'\sim 1$) leads to a stronger condition than 
\eq{luvbound}.\footnote
{In advocating the condition $M\sim \luv$
we differ from Randall et al.~\cite{rsg},
who 
advocate $M\sim\mpl$ even in the presence of the ultra-violet cutoff.}
With $n=4$ one has 
\be
\luv\gsim 10^{12}\GeV \,.
\ee

Finally we mention a different generalization of the original model
\cite{steve}, which is to introduce an interaction
$-A\phi\chi^2$ which dominates the term $-\frac12\mpsis\chi^2$ 
\cite{steve}. This amounts to the replacement $\mpsis\to A^2/\lambda'$,
which does not affect the main result \eq{fourth}.

\paragraph{Inflation at the TeV scale}
Although hybrid inflation remains at least as promising as the 
alternative of single-field inflation, there are significant constraints on the
parameter space. 
As an important example, we consider recent proposals 
\cite{juan,kl,halyo244,halyo223} for hybrid inflation
at and below the TeV scale ($V_0^{1/4} \sim 1\TeV$). 

In \cite{juan} an era of TeV-scale inflation is 
invoked to provide a new model of baryogenesis, in which the Higgs
field is created by the oscillation of the inflaton field.
To achieve this, the trigger field is identified with the
electroweak Higgs field which requires
$\lambda\sim 1$ in \eq{fullpot},
corresponding to $M\sim \TeV$. The loop correction is ignored to produce 
an apparently viable tree-level model with $\phi\sub{COBE}\sim \TeV$,
satisfying the COBE normalization.
But \eq{fourth} shows that with the loop correction included,
 any COBE-normalized model in which the Higgs is the trigger field will 
have $\phi\sub{COBE}\gsim
\phi\sub c\gsim
10^{33}\GeV$, a value presumably too large to contemplate.
Even if we abandon the COBE normalization by assuming
that the TeV-scale inflation lasts only a few 
$e$-folds, \eq{43} requires $\phi\sub{COBE}\sim \mpl$. With such a value
non-renormalizable terms are likely to spoil inflation, and even if they 
were ignored one would have to see if this completely
different model of baryogenesis is still viable. 
Alternatively, one might explore the possibility that the Higgs couples
to the inflaton in a non-hybrid model, still taking on board the fact 
that the coupling to the Higgs will induce a loop correction to the 
inflaton potential. To summarize briefly, the 
specific model of baryogenesis written down in \cite{juan} is 
invalidated by the loop correction, and it remains to be seen whether
alternatives in the same spirit can be constructed.

The model we just looked at assumes that quantum gravity is at the usual
scale $\mpl$. The rest of this note deals with the possibility that it 
is much lower, focussing on the lowest conceivable scale $\luv\sim
\TeV$.
Kaloper and Linde \cite{kl} explore the
possibility of COBE-normalized hybrid inflation in this case.
They invoke the
COBE-normalized tree-level model with $\lambda\sim
\lambda'\sim 1$, corresponding to $m_\chi\sim M\sim \phi\sub{COBE}
\sim V_0^{1/4}\sim \TeV$ and (because of the COBE normalization)
$m\lsim 10^{-7} H$. They point out that the low value of $m$ 
makes the model rather unattractive, because 
supergravity will typically generate $m$ at least of order $H$.
We have seen that with the above parameter choice the loop correction
will invalidate the tree-level model, which means that the above 
difficulty does not in fact exist; the bound $m\lsim 10^{-7} H$ does not 
in fact apply. However, 
\eq{fourth} points to another problem; {\em any} COBE-normalized
model of hybrid inflation 
with $\luv\sim \TeV$ is problematic because
when the loop correction is included the inflaton and/or the trigger field 
have to be at least of order $10^{10}\GeV$.
The bottom line here is that we concur with the conclusion of
Kaloper and Linde, that hybrid inflation with quantum gravity at the
TeV scale looks very problematic.

As
mentioned earlier, Halyo
\cite{halyo244} 
takes the view that field values $\gg\TeV$ are capable of being 
justified, by making the trigger field a superstring modulus.
He advocates
\cite{halyo244} the usual $D$-term hybrid inflation 
model ($\Delta V$ given by \eq{sp1}), with
the quantum gravity scale of order $\TeV$, 
 and with a very small coupling
$\lambda
\sim 10^{-27}$
derived from a superstring theory with large extra dimensions.
In apparent contradiction with our general bound \eq{fourth},
he concludes that an `initial' inflaton field value
(defined as the value when $N=4\times 10^5$) and $M$ can both be of order
$10^6\GeV$. This conflict can be traced to the fact that
Halyo uses
a wrong formula for the 
COBE normalization.
On the reasonable assumption $\phi\sub{COBE}\gg\phi\sub c$, the
correct formula \cite{giaf}
is 
\be
V^{1/4}= \(\frac{50}{NC} \)^{1/4} \(\frac\lambda 4 \)^{1/4}
\times 6 \times 10^{15}\GeV \label{correct}
\,,
\ee
where $C$ is the number of charged pairs coupling to 
the inflaton.\footnote
{Note that the $g$ in Eq.~(246) of \cite{treview} should
actually be $g^{1/2}$.}
However, this formula confirms the
basic tenet of \cite{halyo244}; that 
with a very small $\lambda$ the $D$-term hybrid inflation model can 
give inflation at the TeV scale, if one accepts
field values much bigger than this scale.

Let us end this discussion by noting that potentials of 
the `new' inflation,
as opposed to hybrid inflation, type
can give COBE-normalized 
TeV-scale inflation with the inflaton field $\lsim \TeV$.
Such potentials are of the form $V=V_0+\Delta V$, with $V_0$ 
dominating and $\Delta V<0$.
It is known that COBE-normalized TeV-scale inflation can be achieved
for the cases $\Delta V\simeq \lambda\phi^4\ln(\phi/\phi_0)$ 
(with $\phi\ll\phi_0$) \cite{kt}, $\Delta V=-c
\luv^{-2}\phi^6+b\luv^{-4}\phi^8$
\cite{bd} and $\Delta V = -\lambda\phi^4+b\luv^{4-q}
\phi^q$ \cite{mytev,treview}.
The middle case invokes no small couplings $(c\sim b\sim 1)$.
The first and third cases require $\lambda\sim 10^{-15}$;
however, as we have seen the first case has to come (in the context of 
global supersymmetry) from a 
non-renormalizable interaction like
$\lambda_6\luv^{-2}\phi^6\chi^2$, leading after soft supersymmetry
breaking to $\lambda = (\lambda_6/8\pi^2)(m_\chi/\luv)^2$, and one 
could have $\lambda_6\sim 1$.

For any kind of slow-roll inflation, 
the inflaton mass {\em during inflation} has 
to be much less than $H=V_0^{1/2}/\mpl$ in order to satisfy the
flatness condition $\eta\ll1$ (barring a fine-tuned accidental 
cancellation between the mass term and the displayed terms).
After inflation it becomes much bigger, both in the above models
and in hybrid inflation, so that there need be no problem with 
reheating.

\paragraph{Further remarks on inflation with quantum gravity at the
TeV scale} 
All of this assumes that the size of the extra dimensions,
responsible for quantum gravity at the TeV scale, have their present 
size while cosmological scales leave the horizon during inflation.
It has been shown \cite{cgt} that 
by the {\em end} of inflation,
they must have their present
size, with an accuracy $10^{-14}(T\sub{RH}/10\MeV)^{3/2}$
where $T\sub{RH}$ is the reheat temperature.
Otherwise,
the moduli corresponding to the extra dimensions
would be overproduced afterwards. To
achieve this accuracy one generally needs
$V^{1/4}\ll \TeV$ by the end of inflation.\footnote
{During inflation the canonically normalized modulus
$\chi$, corresponding to the size of the extra dimensions,
is displaced from its minimum by \cite{cgt}
$\delta\chi=\sqrt{n/(n+2)}2H^2\mpl/m_\chi^2$,
where $m_\chi$ is its mass and $n$ is the number of extra dimensions.
This agrees with a rough earlier estimate \cite{thermal}
$\delta\chi\sim H^2\mpl/m_\chi^2$, made for moduli in general.}
(The upper bound on $V^{1/4}$ depends on the number of extra dimensions 
and the moduli mass.)

While cosmological scales are leaving the horizon during inflation,
the extra dimensions must be stabilized, since
significant variation would spoil the observed scale independence of the
spectrum of the primordial curvature perturbation.
The simplest hypothesis is that they remain stabilized 
thereafter, so that they indeed have their present value while cosmological
scales leave the horizon. 

An alternative  \cite{kl,adkm} is to assume that the extra dimensions
are stabilized, while
cosmological scales 
are leaving the horizon, 
with sizes much smaller than at present. In that case, 
after going to the four-dimensional Einstein frame,
the scale of quantum gravity will be far above the TeV scale
and it becomes easier to construct a COBE-normalized model of inflation.
This hypothesis of `asymmetric' inflation has the disadvantage of 
invoking two separate epochs of stabilization for the extra dimensions. 
(It should be carefully distinguished from the idea 
that the extra dimensions must be very small at the beginning of 
inflation, long before cosmological scales leave the horizon.
This indeed seems to be desirable on quite general grounds \cite{kl}.)

Irrespective of the asymmetric inflation hypothesis, or indeed of the 
scale of quantum gravity, one may need a few $e$-folds of
late inflation to get rid of unwanted relics (including moduli)
produced after the 
slow-roll inflation that generates structure.
Suppose first that the late inflation
is hybrid inflation. 
The COBE constraint \eq{fourth} on a hybrid inflation 
model disappears,
but \eq{fifth} remains. Assuming
that all field values should be $\lsim\luv$, this requires
$\luv\lsim 10^4 N^{1/3}\GeV$. A hybrid inflation model
saturating this bound can provide a few $e$-folds of inflation
with quantum gravity at the scale $\luv\sim 10\TeV$.
It is clear that a model saturating the bound would have
$\Delta V$ given by \eq{sp1}, corresponding to
spontaneously broken global susy with supergravity 
corrections assumed to be negligible. 
The usual $D$-term model is of this kind, and 
indeed Halyo
has already noted \cite{halyo223} 
the fact that a few $e$-folds of such inflation is possible,
with both quantum gravity and the field values 
at the $10\TeV$ scale.
One should however remember that even in the $D$-term
(as opposed to an $F$-term \cite{giaf}) model,
non-renormalizable terms in the superpotential may be a 
problem if $\phi$ is at the scale $\luv$ \cite{km,treview}.
What one really needs to be comfortable is $\phi\ll \luv$,
which would require $\luv\gg 10^4\GeV$.

Instead of hybrid inflation one may consider a `new inflation'
potential or some other slow-roll potential \cite{late}.
A completely different alternative  is
thermal inflation \cite{thermal}.
In this paradigm, 
the zero-temperature inflaton potential is $V_0-\frac12m^2\phi^2$,
with $|m|\ll V_0^{1/4}$, but unsuppressed couplings lead to 
a thermal correction  of order $T^4+T^2\phi^2$. In the regime
$m\lsim T\lsim V_0^{1/4}$, some $N\sim \ln(V_0^{1/4}/m)$
$e$-folds of thermal inflation can occur, with $\phi$ held at the origin,
after which $\phi$ reverts to its vev 
\be
\langle\phi\rangle \sim V_0^{1/2}/m \,.
\ee
The expected decay  time of the inflaton
is \cite{thermal}
$\Gamma^{-1} \sim 10^2 \langle\phi\rangle^2/m^3$.
This corresponds to a reheat temperature
\be
T\sub{RH} \sim \(\frac{\TeV}{\langle\phi\rangle}\)^\frac 52
\(\frac{V_0^\frac14}{100\MeV}\)^3 \times 3\MeV \,.
\ee
This can satisfy the cosmological constraints, including the
nucleosynthesis requirement
$T\sub{RH}> \MeV$.
We see that 
thermal inflation, originally proposed in the context of quantum
gravity at the usual scale $\mpl$,
may be viable also in the context of $\TeV$-scale
gravity.\footnote
{It has been invoked in this context
by Dvali \cite{dvalit} in the special case
that the inflaton 
corresponds to the separation between $D$-branes, but the above
discussion following the original papers 
shows the required properties of the inflaton are model-independent, and 
in no way special.}
The main difference from the usual case 
is that 
the required mass $m$ has no obvious explanation.
(In the usual
case one invokes a mass $\sim 100\GeV$, which is natural
in the context of gravity-mediated supersymmetry breaking.)

We end by revisiting an earlier paper \cite{mytev}, which concluded
that TeV-scale quantum gravity does not avoid the need for 
supersymmetry.
This conclusion was based on the fact that 
the inflaton mass during slow-roll inflation has to be
much less than $H\simeq V_0^{1/2}/\mpl\ll V_0^{1/4} \lsim \luv$.
We would like to point that some apparent
escape routes from that conclusion can now be closed off.
The first of these is the possibility 
that the extra dimensions during inflation might be much 
smaller than at present \cite{mytev,kl,adkm}.
As we seen, explicit calculation \cite{adkm,cgt}
has shown
that this cannot be the case during, at least, the last few
$e$-folds. The second is that the last few $e$-folds of inflation
might be thermal as opposed to slow-roll. As we have seen,
this too would require an inflaton mass orders of magnitude
below the TeV scale.

Another possibility might be to have slow-roll inflation, with all
couplings of the inflaton very small. Then the non-supersymmetric
radiative correction $\delta m\sim g\sub{max} \luv$ might be small
enough without any need of supersymmetry.
In that case \cite{no}, reheating could take place only
by the gravitational production of particles, with mass
much less than $H\sim V_0^{1/2}/\mpl$ 
and unsuppressed couplings.\footnote
{We discount here the possibility that
the couplings of the inflaton could somehow be suppressed during
inflation and unsuppressed afterwards. There does not seem to be 
any known way of achieving this.}
Such particles would require supersymmetry to protect their masses.

Barring unforseen escape routes, it seems that even
with quantum gravity at the TeV scale,
one needs during (at least) the last few $e$-folds of inflation
a very light scalar particle, whose mass will have to be protected
by supersymmetry. This means \cite{mytev}, contrary
to what was initially hoped \cite{add},
that there is no particular reason why Nature
should 
have placed the scale of quantum gravity at the TeV scale, as 
opposed to somewhere else in 
the range $\TeV\lsim \luv\lsim \mpl$.
Indeed, while the TeV choice removes
the need for supersymmetry in the Higgs sector of the theory
\cite{add}, the need reappears in the inflaton sector.
The conclusion is that we are unlikely to observe quantum
gravity at a future collider, though of course we may be lucky!

\section*{Acknowledgements}
I am indebted to Toni Riotto, Andrei Linde and Andre' Lukas
for useful discussions.

\newcommand\pl[3]{Phys. Lett. #1 (19#3) #2}
\newcommand\np[3]{Nucl. Phys. #1 (19#3) #2}
\newcommand\pr[3]{Phys. Rep. #1 (19#3) #2}
\newcommand\prl[3]{Phys. Rev. Lett. #1 (19#3) #2}
\newcommand\prd[3]{Phys. Rev. D #1 (19#3) #2}
\newcommand\ptp[3]{Prog. Theor. Phys. #1 (19#3) #2}

\end{document}